\documentclass[12pt]{article}

\usepackage{graphicx}
\usepackage{latexsym}
\usepackage{epsf}
\usepackage{epsfig}
\usepackage{amssymb,amsmath}
\usepackage[english]{babel}
\usepackage{amsfonts}
\usepackage{amssymb}
\usepackage{graphics}
\usepackage{mathrsfs}
\usepackage{verbatim}
\usepackage{float}
\usepackage{slashed}
\usepackage {tabularx}
\newcommand{\nc}{\newcommand}

\nc{\beq}{\begin{equation}}
\nc{\eeq}{\end{equation}}
\nc{\beqa}{\begin{eqnarray}}
\nc{\eeqa}{\end{eqnarray}}
\nc{\bea}{\begin{eqnarray}}
\nc{\eea}{\end{eqnarray}}
\nc{\ra}{\rightarrow}
\nc{\lsim}{\begin{array}{c}\,\sim\vspace{-21pt}\\< \end{array}}
\nc{\gsim}{\begin{array}{c}\sim\vspace{-21pt}\\> \end{array}}

\title{
\vspace*{5mm} \Large\textbf{On Taming the Warped Radion with Supersymmetry}
\vspace*{1.0cm}
\author{\textbf{Hooman Davoudiasl~$^a$ and Eduardo Pont\'on~$^b$}\\\\
\normalsize\emph{$^a$Department of Physics, Brookhaven National Laboratory,
Upton, NY 11973, USA}\\
\normalsize\emph{$^b$Department of Physics, Columbia University,
538 W. 120th St, New York, NY 10027, USA}}}
\date{\today}
\begin{document}
\setcounter{page}{0}
\maketitle
\begin{abstract}

In warped models that solve the hierarchy problem, there is generally
no dynamical relation between the size of the fifth dimension and the
scale of electroweak symmetry breaking (EWSB).  The establishment of
such a relation, without fine-tuning, requires that Casimir
contributions to the radion potential not exceed the energy density
associated with EWSB. Here, we examine the use of supersymmetry for
controlling the Casimir energy density and making quantum
contributions calculable.  We compute the effects of supersymmetry
breaking at the UV and IR boundaries of warped backgrounds, in the
presence of brane localized kinetic terms.  Various limits of
supersymmetry breaking are examined.  We find that when supersymmetry
is broken on the UV brane, vacuum contributions to the radion
potential can be controlled (as likely necessary for EWSB to govern
the radion potential) via small soft masses as well as a ``double
volume suppression."  Our formalism can also provide a setup for
radion stabilization by bulk fields, when supersymmetry is broken on
both the UV and the IR branes.

\end{abstract}

\thispagestyle{empty}
\newpage
\setcounter{page}{1}

\section{Introduction}
\label{sec:Intro}

An interesting resolution of the gauge hierarchy problem is offered by
warped models based on the original Randall-Sundrum (RS) geometry
\cite{RS} which is a slice of the 5D Anti de Sitter (AdS$_5$)
spacetime.  This geometry, characterized by a curvature scale $k$, is
bounded by flat walls, often referred to as the UV and IR branes.  In
these models the fundamental 5D scales of the theory, perhaps near the
4D Planck scale $M_P \sim 10^{19}$~GeV, redshift to the weak scale of
order 1~TeV as one goes from the UV brane to the IR brane.  The
redshift is caused by an exponential warp factor and a Planck-weak
hierarchy of order $10^{16}$ can be achieved if the size $L$ of the
extra dimension satisfies $k L \gtrsim 30$.  In addition, placing the
Standard Model (SM) gauge \cite{Davoudiasl:1999tf,Pomarol:1999ad} and
fermion \cite{Grossman:1999ra} content in all 5D can yield a realistic
explanation of 4D flavor \cite{Gherghetta:2000qt}, where heavy
fermions are localized towards the source of the electroweak symmetry
breaking (EWSB) on the IR brane, while light fermions are UV-brane
localized, so that their small masses arise from an exponentially
small wavefunction overlap with the IR brane.

While many features of the SM can be accommodated and explained in
warped scenarios, quite often the scale $m_W \sim 10^2$~GeV of EWSB is
simply assumed to be set by the IR-brane physical scale
$\tilde{k}\equiv k \, e^{- k L}$.  Since various bounds from precision
and collider phenomenology favor $\tilde{k}\gtrsim 1$~TeV, this often
reintroduces some level of fine-tuning into the models.  It would then
be interesting to find a dynamical mechanism whereby EWSB sets $kL$
near the requisite value, and that naturally leads to a separation
between $m_{W}$ and $\tilde{k}$.  As the size of the extra dimension
is controlled by the dynamics of the radion scalar, one would like to
generate an appropriate potential for the radion through EWSB.

In fact, such a model was proposed in Ref.~\cite{Bai:2008gm}, where
the condensation of IR-localized top quarks of the SM is triggered by
strong interactions mediated by Kaluza-Klein (KK) gluons.  This
condensation breaks the electroweak symmetry and through its
radion-dependence can also stabilize the size of the fifth dimension.
In this top-condensation scenario, the IR-bane scale is set by
$\tilde{k}\sim 10$~TeV, which has the added benefit of avoiding most
severe constraints on warped models, while providing a GeV-scale
radion as its low energy signature
\cite{Bai:2008gm,Davoudiasl:2009xz}.  Due to the dynamical nature of
the condensation, the separation between the IR-brane scale and the
weak scale is natural.  However, one has to assume that other
contributions to the radion potential would not overwhelm the
electroweak contribution.  In particular, one may naively expect that
the corrections from zero-point energies of bulk fields to the radion
potential (which contain the calculable Casimir energy) are of order
$\tilde{\Lambda}^4/(16 \pi^2)$, where the effective theory
cutoff-scale $\tilde{\Lambda} \sim 4\pi \tilde{k}$.

Given the above situation, either one has to resort to a severe
fine-tuning, or else find a way to suppress such cutoff scale effects.
In this work, we adopt the second possibility and examine how
supersymmetry (SUSY) can be used to protect the radion potential from
too large cutoff dominated contributions.  Since in the limit of exact
SUSY all vacuum energies cancel out, in supersymmetric scenarios the
full radion potential is expected to be far less sensitive to the
cutoff.  We will consider soft supersymmetry breaking by localized
mass terms on both the UV and the IR branes.  In addition, we allow
for the presence of brane kinetic terms (BKTs), which are expected to
be induced radiatively, and that are known to have a strong effect on
the KK spectrum~\cite{Davoudiasl:2002ua, Carena:2002dz,
Carena:2004zn}.  We find that when SUSY is broken on the UV brane,
this vacuum energy contribution is of order $\tilde{k}^{2}
m_{\lambda}^{2}/(16\pi^{2} kL)$, where $m_{\lambda}$ is the physical
mass of the (would-be zero-mode) gaugino.  Hence, there is a
suppression from the dependence on soft masses, as well as an
additional volume-suppression.  This result suggests that a
supersymmetric version of the scenario presented in
Ref.~\cite{Bai:2008gm} may be realizable without the need for
fine-tuning.  Even though dynamical EWSB scenarios, which will be
pursued in Ref.~\cite{HDEP}, provide our basic motivation for this
study, we will also consider the effect of a supersymmetric bulk in
other contexts.  In particular, in models where the Casimir energy is
responsible for generating the radion potential
\cite{Garriga:2000jb,Goldberger:2000dv,Garriga:2002vf,Maru:2010ap},
the success of the mechanism depends on the size and sign of certain
contributions that are in general not calculable.  We show how a
supersymmetric framework with SUSY broken at a high scale may realize
the radion stabilization mechanism by gauge fields proposed in
Ref.~\cite{Garriga:2002vf}.  The above radion stabilization mechanisms
are distinct from stabilization by bulk fields that acquire vacuum
expectation values (VEV's) at tree level, as in
Ref.~\cite{Goldberger:1999uk}.  Related work has appeared
in Refs.~\cite{Flachi:2003bb, Gregoire:2004nn, Katz:2006mva}.

The structure of this paper is as follows.  In the next section, we
consider the case of a bulk scalar field, with brane localized masses
and kinetic terms, and evaluate the radion potential generated by the
Casimir effect from the scalar.  The results are sufficiently general
to be applied to fields of arbitrary spin that obey arbitrary boundary
conditions.  We provide such a generalization to particles of
different spin in Section~\ref{sec:OtherSpins}.  We apply our results
to study scenarios in which supersymmetry is broken on the UV or IR
branes, as may be relevant in different models, in
Sections~\ref{sec:susy}.  We present our conclusions in
Section~\ref{sec:conclusions}, followed by
Appendices~\ref{app:scalars} (conventions for KK reduction) and
\ref{app:AnalyticApprox} (approximate expressions for the radion
potential in various limits).

\section{Radion Potential from Scalar Fields}
\label{sec:RadionPotential}

In this section, we summarize the results for the radion potential at
one-loop order in warped backgrounds.  Since this potential arises
through the radion-dependence of the KK masses associated with bulk
fields propagating in compact dimensions, we focus on describing the
spectrum of such fields.  The spectrum depends on the quadratic terms
in the action (including both bulk and brane-localized operators), as
well as on the boundary conditions.  Since the results for fields in
different spin representations of the Lorentz group can be expressed
in terms of those of a scalar field, we start with the real scalar case.
For reference, further details regarding the boundary conditions, the
KK decomposition, orthonormality relations, etc.  are relegated to
Appendix~\ref{app:scalars}.

\subsection{Bulk Fields with Brane-Localized Terms}
\label{sec:scalars}

We consider a slice of AdS$_{5}$:
\beqa
ds^2 = e^{-2ky} \eta_{\mu\nu} dx^{\mu} dx^{\nu} - dy^{2}~,
\label{metric}
\eeqa
where $k$ is the AdS curvature and $y \in [0,L]$ parametrizes the
fifth dimension.  The action for a free real scalar field is
\beqa
S = \int \! d^{4}x \int_{0}^{L} \! dy \, \sqrt{g}
\left\{ \frac{1}{2} g^{MN} \partial_{M} \Phi \partial_{N} \Phi -
\frac{1}{2} M^{2} \Phi^{2}
+ 2\delta(y) {\cal L}_{0} +
2\delta(y-L) {\cal L}_{L} \right\}~,
\label{Action}
\eeqa
where the brane-localized terms are given by~\footnote{For simplicity,
we restrict to brane kinetic terms involving only $\mu$ derivatives,
but the same methods can be easily generalized to include arbitrary
BKTs by using the appropriate eigenvalue equation.  This may require
an appropriate classical renormalization when $\partial_{5}$
derivatives are involved~\cite{delAguila:2003bh}.}
\beqa
{\cal L}_{0} &=& \frac{1}{2} r_{UV} \, g^{\mu\nu} \partial_{\mu} \Phi \partial_{\nu} \Phi - \frac{1}{2} M_{0} \Phi^{2}~,
\\[0.5em]
{\cal L}_{L} &=&  \frac{1}{2} r_{IR} \, g^{\mu\nu} \partial_{\mu} \Phi \partial_{\nu} \Phi - \frac{1}{2} M_{L} \Phi^{2}~.
\eeqa
The kinetic coefficients $r_{UV}$ and $r_{IR}$ have mass
dimension~$-1$, while $M_{0}$ and $M_{L}$ have mass dimension~$1$.  It
is convenient to introduce a dimensionless parameter $\alpha$ (that we
will call the ``index'' of the field) and new mass parameters $m_{UV}$
and $m_{IR}$ by writing the bulk and localized masses as
\beqa
M^{2} &=& \left(\alpha^{2} - 4 \right) k^{2}~,
\label{Mdef}
\eeqa
\vspace{-7mm}
\beqa
M_{0} &=& - \left(\alpha - 2 \right) k + m_{UV} ~,
\label{m0def} 
\hspace{1cm}
M_{L} ~=~ \left(\alpha - 2 \right) k + m_{IR} ~.
\label{mLdef}
\eeqa
Performing the KK decomposition in the usual manner (see
Appendix~\ref{app:scalars}), one finds that the KK spectrum, $m_{n}$,
is determined by~\footnote{$J_{\alpha}$ and $Y_{\alpha}$
($I_{\alpha}$ and $K_{\alpha}$) are the (modified) Bessel functions of
the first and second kind, respectively.}
\beqa
F(m_{n}/k) \equiv \tilde{J}^{IR}_{\alpha} \! \left(\frac{m_{n}}{k} e^{kL} \right)
\tilde{Y}^{UV}_{\alpha} \! \left(\frac{m_{n}}{k} \right)
- \tilde{J}^{UV}_{\alpha} \! \left(\frac{m_{n}}{k} \right)
\tilde{Y}^{IR}_{\alpha} \! \left(\frac{m_{n}}{k} e^{kL}  \right) &=& 0~,
\label{Feigen}
\eeqa
where $L$ is the radion VEV.  Here, we have introduced the functions
\beqa
\tilde{J}^{UV}_{\alpha}(z) &=& z J_{\alpha-1}(z) + b_{UV}(z) J_{\alpha}(z)~,
\label{JUVtilde}
\\ [0.5em]
\tilde{J}^{IR}_{\alpha}(z) &=& z J_{\alpha-1}(z) - b_{IR}(z) J_{\alpha}(z)~,
\label{JIRtilde}
\eeqa
with analogous expressions for $\tilde{Y}^{UV}_{\alpha}(z)$ and
$\tilde{Y}^{IR}_{\alpha}(z)$, and defined
\beqa
b_{i}(z) &=& \hat{r}_{i} z^{2} - \hat{m}_{i}~,
\hspace{1cm}
i = UV, IR
\label{bUVIR}
\eeqa
by using the dimensionless parameters $\hat{m}_{i} = m_{i}/k$,
$\hat{r}_{i} = k \, r_{i}$.

For $\hat{m}_{UV} = \hat{m}_{IR} = 0$ (the SUSY limit) there is a
scalar zero-mode with an exponential profile controlled by the
dimensionless parameter $\alpha$ (see Appendix~\ref{app:LightStates}
for details).  This profile coincides with that of a 5D fermion with a
bulk Dirac mass term written as $M = c k$, where $\alpha = c +
\frac{1}{2}$~\cite{Grossman:1999ra}.  We will refer to fields with
$\alpha < 1$ as ``IR localized fields'', those with $\alpha = 1$ as
``flat fields'' and those with $\alpha > 1$ as ``UV localized
fields''.  When $\hat{m}_{UV}$ and/or $\hat{m}_{IR}$ deviate from
zero, the scalar zero-mode is lifted: for positive $\hat{m}_{UV, IR}$
it becomes massive, while for negative $\hat{m}_{UV, IR}$ a tachyonic
mode appears.  Note also that one can interpolate between Neumann-type
boundary conditions (setting $m_{i} = 0$) and Dirichlet boundary
conditions ($m_{i} \to \infty$) on the $i^{\rm th}$ ($=UV$ or $IR$)
brane, and therefore our results are sufficiently general to capture
the dependence of the radion potential on the choice of boundary
conditions for the bulk field.

It is also worth pointing out that, when $\hat{r}_{i} = \hat{m}_{i} =
0$, the massive KK spectrum is identical for fields with index $\alpha
= 1 + \delta\alpha$ and fields with index $\alpha = 1 - \delta\alpha$,
for any $\delta\alpha$, as can be easily checked from
Eq.~(\ref{Feigen}).  However, non-vanishing $\hat{r}_{i}$ and/or
$\hat{m}_{i}$ introduce a distinction between the KK spectra of UV
versus IR localized fields.

\subsection{One-Loop Effective Potential}
\label{sec:RadionPotentialResults}

The presence of fields propagating in a $D$-dimensional spacetime with
$D-4$ compact dimensions generically leads to a vacuum energy that
depends on the overall volume of the extra dimensions as well as on
the various shape moduli.  For a single compact extra dimension one
simply gets a potential for its size $L$, given at one-loop order by
\beqa
V_{\rm eff}(L) &=& \frac{1}{2} \sum_{n} \int \! \frac{d^{4} p_{E}}{(2\pi)^{4}} \, \ln\left( \frac{p^{2}_{E} + m^{2}_{n}}{\mu^2} \right)~,
\label{EffPot}
\eeqa
where the radion dependence enters through the spectrum, $m_{n}(L)$.
Here the integration is over Euclidean momentum and $\mu$ is the
renormalization scale.  Eq.~(\ref{EffPot}) is the contribution due to
a bulk real scalar, but we will consider other types of fields in
Section~\ref{sec:OtherSpins}.  In this work, we will assume that the
backreaction due to the one-loop vacuum energy is small, so that the
AdS$_{5}$ metric of Eq.~(\ref{metric}) gives a good approximation.
The effective potential contains divergent pieces that correspond to a
renormalization of the 5D cosmological constant, as well as of the IR
and UV brane tensions.  The first two lead to radion-dependent terms
that behave like $e^{-4kL}$.  Additionally, there are calculable
terms, usually referred to as the Casimir energy, that can have a more
complicated radion dependence and can sometimes stabilize the distance
between the branes~\cite{Garriga:2002vf}.  In order to exhibit the
various radion-dependent contributions, one can evaluate
Eq.~(\ref{EffPot}) using standard methods based on $\zeta$-function
regularization.  The \textit{regularized} radion potential has been
computed in
Refs.~\cite{Garriga:2000jb,Goldberger:2000dv,Garriga:2002vf,Maru:2010ap},
and takes the form
\beqa
V_{\rm eff}(L) &=& \frac{k^{4}}{16\pi^{2}} \left[  {\cal I}_{UV} + e^{-4kL} {\cal I}_{IR}
\right] + V_{\rm Casimir}(L)~,
\label{FinalPot}
\eeqa
where the ``Casimir energy'' is given by~\footnote{Throughout this
work, the designation {\it calculable} refers to effects that are
insensitive to the physics near or above the cutoff of the 5D
effective theory.  Here, we will refer to Eq.~(\ref{CasimirPot}) as
the ``Casimir energy", even though it can contain terms that scale
like $e^{-4kL}$, exactly as the IR brane tension contribution, and
that should not be considered calculable in generic theories.}
\beqa
V_{\rm Casimir}(L) &=& \frac{k^{4} e^{-4kL}}{16\pi^{2}} \int_{0}^{\infty} \! dt \, t^{3} \ln \left| 1 - \frac{\tilde{K}^{IR}_{\alpha}(t)}{\tilde{I}^{IR}_{\alpha}(t)} \frac{\tilde{I}^{UV}_{\alpha}(t e^{-kL})}{\tilde{K}^{UV}_{\alpha}(t e^{-kL})} \right|~,
\label{CasimirPot}
\eeqa
and the ``generalized'' Bessel functions are defined by
\beqa
\tilde{K}^{UV}_{\alpha}(z) &=& z K_{\alpha-1}(z) - \tilde{b}_{UV}(z) K_{\alpha}(z)~,
\label{tildeKUV}
\\ [0.5em]
\tilde{I}^{UV}_{\alpha}(z) &=& z I_{\alpha-1}(z) + \tilde{b}_{UV}(z) I_{\alpha}(z)~,
\label{tildeIUV}
\\ [0.5em]
\tilde{K}^{IR}_{\alpha}(z) &=& z K_{\alpha-1}(z) + \tilde{b}_{IR}(z) K_{\alpha}(z)~,
\label{tildeKIR}
\\ [0.5em]
\tilde{I}^{IR}_{\alpha}(z) &=& z I_{\alpha-1}(z) - \tilde{b}_{IR}(z) I_{\alpha}(z)~.
\label{tildeIIR}
\eeqa
Here, $\tilde{b}_{i}(z) \equiv - \hat{r}_{i} z^{2} - \hat{m}_{i}$ has
an additional minus sign in the first term compared to
Eq.~(\ref{bUVIR}) as a result of the Wick rotation involved in
obtaining Eq.~(\ref{FinalPot}).

The terms proportional to ${\cal I}_{UV}$ and ${\cal I}_{IR}$ in
Eq.~(\ref{FinalPot}) correspond to UV and IR brane tension
renormalizations, respectively.  (${\cal I}_{UV}$ depends only on the
UV quantities $\hat{m}_{UV}$ and $\hat{r}_{UV}$, while ${\cal I}_{IR}$
depends only on the IR quantities $\hat{m}_{IR}$ and $\hat{r}_{IR}$.)
Such contributions are in general uninteresting, since there are
additional incalculable (and most likely larger) contributions that
renormalize the brane tensions.  Hence, we will focus on the Casimir
term in what follows.

\subsection{The Casimir Energy}
\label{CasimirEnergy}

The Casimir energy contribution to the radion potential,
Eq.~(\ref{CasimirPot}), which contains the 5D non-local, and therefore
calculable, terms has been computed in the
literature~\cite{Garriga:2000jb,Goldberger:2000dv,Garriga:2002vf}.
Although the expression given in Eq.~(\ref{CasimirPot}) can be easily
evaluated numerically for any value of $L$, and for fixed values of
the parameters in the Lagrangian: $\alpha = c+1/2$, $m_{i}$ and
$r_{i}$ ($i=UV$ or $IR$), it is useful to have analytic approximations
that make the radion dependence more explicit.  We derive general
expressions for the case that $e^{-k L} \ll 1$ in
Appendix~\ref{app:AnalyticApprox}.  These result from the fact that
the radion dependence is fully contained in the ``UV factor''
$\tilde{I}^{UV}_{\alpha}(t e^{-kL})/\tilde{K}^{UV}_{\alpha}(t
e^{-kL})$, which can legitimately be approximated by a small argument
expansion of the Bessel functions.  Thus, the functional dependence of
the Casimir energy on $L$ is determined by the UV quantities.  In
particular, one finds that $\hat{m}_{UV}$ defines a characteristic
scale $L_{T}$ according to~\footnote{For $\alpha < 0$, one has $kL_{T}
\ll 1$, unless $\hat{m}_{UV}$ is exponentially large.}
\beqa
\begin{array}{lcl}
e^{2kL_{\rm T}}\hat{m}_{UV} = 1~,
  &   & \textrm{for } \alpha \geq 1~,  \\ [0.2em]
  & \textrm {or} & \\ [0.2em]
e^{2\alpha kL_{\rm T}}\hat{m}_{UV} = 1~,
  &   & \textrm{for } \alpha < 1~.
\end{array}
\label{LT}
\eeqa
As shown in Appendix~\ref{app:LightStates}, the above transition
signals whether a KK state parametrically lighter than the KK
scale, $k \, e^{-kL}$, exists or not.  For $L \gg L_{T}$ (``large
$m_{UV}$'') and $\alpha > 0$, one finds
\beqa
V_{\rm Casimir} &\approx& \frac{k^{4} e^{-2(2+\alpha)kL}}{16\pi^{2}} \frac{2^{1-2\alpha}(\hat{m}_{UV} - 2 \alpha) }{\hat{m}_{UV} \Gamma(\alpha) \Gamma(\alpha + 1)}
\int_{0}^{\infty} \! dt \, t^{3+2\alpha} \frac{\tilde{K}^{IR}_{\alpha}(t)}{\tilde{I}^{IR}_{\alpha}(t)}~,
\label{VCasLargemUV}
\eeqa
while the expressions for $\alpha = 0$ and $\alpha < 0$ are given in
Eqs.~(\ref{VCasimirApproxLargemUValpha0}) and
(\ref{VCasLargemUVNegalpha}) of Appendix~\ref{app:AnalyticApprox},
respectively.  For $L \ll L_{T}$ (``small $m_{UV}$''), which is the
case with a light would-be zero mode in the spectrum, but $kL \gg 1$,
one has to distinguish several cases.  For instance, for moderate
localization of this mode near the IR brane, one finds
\beqa
V_{\rm Casimir} &\approx& \frac{k^{4} e^{-4kL}}{16\pi^{2}} 
\int_{0}^{\infty} \! dt \, t^{3} \ln\left| 1 - \frac{2 \sin(\pi\alpha)}{\pi} \, \frac{\tilde{K}^{IR}_{\alpha}(t)}{\tilde{I}^{IR}_{\alpha}(t)} \right|~,
\hspace{1cm} \textrm{for } 0 < \alpha < 1~.
\label{VCasSmallmUV0alpha1}
\eeqa
For the flat case, $\alpha = 1$:
\beqa
V_{\rm Casimir} &\approx& - \frac{k^{4} e^{-4kL}}{16\pi^{2}} \, \frac{1}{k L + \hat{r}_{UV}}
\int_{0}^{\infty} \! dt \, t^{3} \frac{\tilde{K}^{IR}_{\alpha=1}(t)}{\tilde{I}^{IR}_{\alpha=1}(t)}~,
\label{VCasSmallmUValpha1}
\eeqa
where, for simplicity, we have ignored the numerically small Euler
constant, as well as a $\ln(t/2)$ term that makes a relatively small
difference [see third line in Eq.~(\ref{UVRatioSmallmUV}) of
Appendix~\ref{app:AnalyticApprox}].  Finally, for moderate
localization near the UV brane one finds
\beqa
V_{\rm Casimir} &\approx& - \frac{k^{4} e^{-2(\alpha+1)kL}}{16\pi^{2}} \, \frac{2^{3 - 2 \alpha}}{\Gamma(\alpha) \left[ \Gamma(\alpha - 1) + 2\hat{r}_{UV} \Gamma(\alpha) \right]}
\int_{0}^{\infty} \! dt \, t^{1+2\alpha} \frac{\tilde{K}^{IR}_{\alpha}(t)}{\tilde{I}^{IR}_{\alpha}(t)}~,
\hspace{5mm} \textrm{for } 1 < \alpha < 2~.
\nonumber \\
\label{VCasSmallmUV1alpha2}
\eeqa
The case of more extreme IR localization (with $\alpha \leq 0$) is
given in Eq.~(\ref{VCasLargemUVNegalpha}), while for more extreme UV
localization (with $\alpha \geq 2$) the result can be easily obtained
from Eq.~(\ref{UVRatioSmallmUV}) of Appendix~\ref{app:AnalyticApprox}.
In Eqs.~(\ref{VCasSmallmUV0alpha1}), (\ref{VCasSmallmUValpha1}) and
(\ref{VCasSmallmUV1alpha2}), we have ignored terms subleading in
$\hat{m}_{UV}$.  In SUSY theories that are softly broken by UV brane
localized masses, the above leading contributions cancel out within a
given supermultiplet, and one should keep the terms subleading in
$\hat{m}_{UV}$.  We postpone such an application to
Section~\ref{sec:susy}.  For the time being we simply make a few
observations that follow from the previous results.  First, when
$\hat{m}_{UV}$ is large in the sense defined above, the Casimir energy
is suppressed by a factor $e^{-2\alpha k L}$ on top of the naive
$e^{-4kL}$ expected from the scaling of the KK masses.  This happens
for any $\alpha > 0$ [see Eq.~(\ref{VCasLargemUV})].  Second, for
small $\hat{m}_{UV}$, fields localized near the IR brane give a radion
dependence identical to the one arising from an IR brane tension, up
to exponentially small terms [see Eq.~(\ref{VCasSmallmUV0alpha1})].
Third, as pointed out in Ref.~\cite{Garriga:2002vf}, flat fields
produce only a volume suppression [see Eq.~(\ref{VCasSmallmUValpha1})]
that allows their contribution to be plausibly comparable to the IR
brane tension contribution, and in certain cases can lead to radion
stabilization.  And finally, fields localized towards the UV brane
give a contribution that is exponentially small compared to that of
the IR brane tension [see Eq.~(\ref{VCasSmallmUV1alpha2})].

\begin{figure}[t]
\begin{center}
\includegraphics[width=0.5 \textwidth]{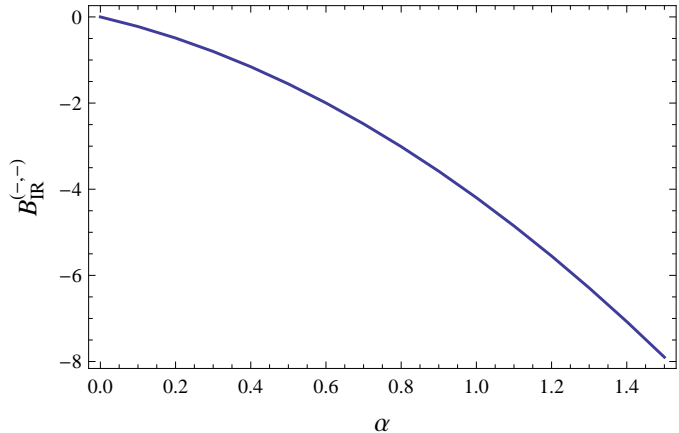}
\caption{$B^{(-,-)}_{IR}$ of Eq.~(\ref{BIR_mmFields_alphaPos}) as a
function of $\alpha$.  This determines the contribution to the
effective potential due to fields obeying $(-,-)$ boundary conditions
according to Eq.~(\ref{VCasminusFields}).  At very small $\alpha$ one
should use Eq.~(\ref{BIR_mmFields_alpha0}), but this detail is
imperceptible in the figure when $kL \gg1$ since $B^{(-,-)}_{IR}$ is
very small.}
\label{fig:BIRvsralpha}
\end{center}
\end{figure}

In applications, one often finds fields obeying Dirichlet boundary
conditions on both branes.  The contribution to the Casimir energy
from such fields can be easily obtained by taking the limits
$\hat{m}_{UV} \to \infty$ and $\hat{m}_{IR} \to \infty$ [e.g.~in
Eq.~(\ref{VCasLargemUV})].  For the Casimir energy part, one finds
that for $\alpha > 0$:
\beqa
V^{(-,-)}_{\rm Casimir} &\approx& \frac{B^{(-,-)}_{IR} k^{4}}{16\pi^{2}} \, e^{-2(2+\alpha)kL}~,
\label{VCasminusFields}
\eeqa
where
\beqa
B^{(-,-)}_{IR} &=& - \frac{2^{1 - 2 \alpha}}{\Gamma(\alpha) \Gamma(\alpha + 1)}
\int_{0}^{\infty} \! dt \, t^{3+2\alpha} \frac{K_{\alpha}(t)}{I_{\alpha}(t)}~
\label{BIR_mmFields_alphaPos}
\eeqa
depends only on the localization of the field.  We plot
$B^{(-,-)}_{IR}$ in Fig.~\ref{fig:BIRvsralpha}, which shows that it is
always negative.  Eq.~(\ref{VCasminusFields}) shows that these effects
are exponentially suppressed compared to the IR brane tension
contribution.  The exception occurs for $\alpha \approx 0$, in which
case one should use [see Eq.~(\ref{VCasimirApproxLargemUValpha0}) of
Appendix~\ref{app:AnalyticApprox}]
\beqa
B^{(-,-)}_{IR} &\approx& - \frac{1}{kL} \int_{0}^{\infty} \! dt \, t^{3} 
\frac{K_{\alpha = 0}(t)}{I_{\alpha = 0}(t)}~,
\label{BIR_mmFields_alpha0}
\eeqa
instead of Eq.~(\ref{BIR_mmFields_alphaPos}), setting $\alpha = 0$ in
Eq.~(\ref{VCasminusFields}). 
For $\alpha < 0$, one finds [see Eq.~(\ref{VCasLargemUVNegalpha}) of
Appendix~\ref{app:AnalyticApprox}, with $\hat{m}_{IR} \to \infty$]:
\beqa
V^{(-,-)}_{\rm Casimir} &\approx& \frac{k^{4} e^{-4kL}}{16\pi^{2}}
\int_{0}^{\infty} \! dt \, t^{3} \ln \left|1 + \frac{2 \sin(\pi\alpha)}{\pi} \, \frac{K_{\alpha}(t)}{I_{\alpha}(t)} \right|~.
\label{BIR_mmFields_alphaNeg}
\eeqa

In fact, in the case of vanishing BKT's, the $(-,-)$ contribution of a
field with index $\alpha$ is identical to that of a field with
vanishing localized masses and index $\alpha+1$ (see
Section~\ref{sec:OtherSpins}).  The results of this section will be
used in Sections~\ref{sec:OtherSpins} and \ref{sec:susy}.

\section{Higher Spins and SUSY Multiplets}
\label{sec:OtherSpins}

In this section we give a brief summary of the contributions to the
radion potential from 5D fields of spin-$1/2$, $1$, $3/2$ and $2$, and
then give the results for the net effective potential from various
SUSY multiplets.  In general, these contributions can be simply
expressed in terms of the effective potential due to a real scalar
field, which was evaluated above.  One finds
\beqa
V_{\rm eff}^{s \geq 0}(L) &=& (-1)^{2s}\, n\, V^{\textrm{Real scalar}}_{\rm eff}(L) ~,
\label{Veff_General}
\eeqa
where $V^{\textrm{Real scalar}}_{\rm eff}(L) = V_{\rm eff}(L)$ is
given in Eq.~(\ref{FinalPot}) and $n$ is the number of physical
degrees of freedom of a \textit{massive} 4D field of spin $s$ (taking
into account whether the field is real or complex, in the bosonic
case).  Deriving the result above is straightforward for complex
scalars or fermions, but requires adding a gauge-fixing term to the
action for higher spins.  One also has to include the associated
Faddeev-Popov determinant, which simply cancels the spurious
contributions to the potential from the gauge degrees of freedom.
Here we just use Eq.~(\ref{Veff_General}) in our applications below
(with $n=2$ for complex scalars, $n=4$ for fermions, $n=3$ for (real)
gauge bosons, $n=5$ for the graviton and $n=8$ for the gravitino).
Based on this, we can put together the total contributions to the
radion potential due to (softly broken) SUSY multiplets.  Since the
compactification breaks the 5D $N=1$ SUSY (4D $N=2$) down to 4D $N=1$
SUSY, it is useful to classify the field content in 4D $N=1$ language.

Starting with hypermultiplets, and listing only the on-shell fields,
these consist of two 4D $N=1$ chiral multiplets: $(\Phi_{1},\psi_{1})$
and $(\Phi_{2},\psi_{2})$.  If the pair $(\Phi_{1},\psi_{1})$ has
``localization parameter'' (or index) $\alpha$, then the second pair
of fields, $(\Phi_{2},\psi_{2})$, has index $\alpha - 1$.  Both
$\Phi_{2}$ and $\psi_{2}$ obey Dirichlet boundary conditions, so that
they do not have zero-modes.  For vanishing kinetic terms,
$\hat{r}_{UV} = \hat{r}_{IR} = 0$, and arbitrary UV and/or IR
localized masses for the complex scalar $\Phi_{1}$ (recall that
$\Phi_{2}$ vanishes on the branes) the effective potential is:
\beqa
\left. \rule{0mm}{5mm} V^{\rm Hyper}_{\rm eff} \right|_{\hat{r}_{i} = 0} 
&=& 2 \times \left. \rule{0mm}{5mm} V_{\rm eff}(L) \right|_{\alpha;\hat{r}_{i} = 0;
\hat{m}_{i}} + 2 \times \left. V^{(-,-)}_{\rm eff}(L) \right|_{\alpha-1} 
- 4 \times \left. \rule{0mm}{5mm} V_{\rm eff}(L) \right|_{\alpha;\hat{r}_{i} = 0;
\hat{m}_{i}=0}~,
\label{Veff_Hyper_nori}
\eeqa
where $V_{\rm eff}(L)$ is given in Eq.~(\ref{FinalPot}), and
$V^{(-,-)}_{\rm eff}(L)$, which depends only on the index $\alpha$, is
the effective potential for a real scalar obeying Dirichlet boundary
conditions on both branes [the (approximate) Casimir energy part is
given in Eqs.~(\ref{BIR_mmFields_alphaPos}),
(\ref{BIR_mmFields_alpha0}) or (\ref{BIR_mmFields_alphaNeg}),
according to the value of $\alpha$].  We have also indicated the index
$\alpha$ of the ``generalized'' Bessel functions [see
Eq.~(\ref{tildeKUV})-(\ref{tildeIIR})] to be used in each term.  The
fermions $\psi_{1}$ and $\psi_{2}$ contribute together as a KK tower
of Dirac fermions through the last term in
Eq.~(\ref{Veff_Hyper_nori}).  The total effective potential above
vanishes in the SUSY limit, i.e.~when $\hat{m}_{i} = 0$.  Indeed, in
this case the approximate expressions~(\ref{VCasSmallmUV0alpha1}),
(\ref{VCasSmallmUValpha1}) or (\ref{VCasSmallmUV1alpha2}) for $V_{\rm
Casimir}$ hold, depending on the value of $\alpha$, and one can
explicitly see that the Casimir energy contributions from the
components of the hypermultiplet cancel out.  More generally, the
exact cancellations follow from
\beqa
\left. \frac{\tilde{K}^{i}_{\alpha}(z)}{\tilde{I}^{i}_{\alpha}(z)} \right|_{\hat{r}_{i} = 0;\hat{m}_{i}=0} &=& 
\left. \frac{\tilde{K}^{i}_{\alpha-1}(z)}{\tilde{I}^{i}_{\alpha-1}(z)} \right|_{\hat{r}_{i} = 0;\hat{m}_{i}=\infty}~,
\hspace{1.5cm}\textrm{for}~i=UV,IR~,
\eeqa
which implies that the Casimir energies, Eq.~(\ref{CasimirPot}), are
the same for a field with index $\alpha-1$ that obeys Dirichlet
boundary conditions (obtained by taking $\hat{m}_{i} \to \infty$), and
for a field with $\hat{m}_{i}=0$ and index $\alpha$ ({\it i.e.}~the KK
spectra are identical in these two cases). 

When the brane kinetic terms are non-zero one has to be more careful,
since supersymmetry requires that the action for the ``odd'' scalar
field, $\Phi_{2}$, contain special singular terms on the branes.
These can be derived by writing the SUSY action in 4D $N=1$ language,
following the formalism of
Refs.~\cite{ArkaniHamed:2001tb,Marti:2001iw}, together with
brane-localized minimal K\"{a}hler terms for the ``even'' chiral
superfield, $(\Phi_{1}, \psi_{1})$, and then integrating out the
auxiliary fields.  Besides brane localized terms (with
$\partial_{\mu}$ derivatives) for $\Phi_{1}$ and $\psi_{1}$, one finds
that the bulk kinetic term for $\Phi_{2}$ should be replaced by
\beqa
|\partial_{M} \Phi_{2}|^{2} \to |\partial_{\mu} \Phi_{2}|^{2} - \frac{1}{1+\sum_{i=UV,IR} r_{i} \delta(y-y_{i})} |\partial_{5} \Phi_{2}|^{2}~,
\label{SingularBKTs_OddScalar}
\eeqa
where $y_{UV} = 0$ and $y_{IR} = L$, while $r_{i}$ are the brane
kinetic coefficients for $\Phi_{1}$ and $\psi_{1}$.  This is
reminiscent of the singular terms first pointed out in
Ref.~\cite{Mirabelli:1997aj}, and implies that the equation of motion
for $\Phi_{2}$ is identical to the second order differential equation
of motion obeyed by $\psi_{2}$.  In particular, the spectrum of
$\Phi_{2}$ at the massive level is identical to the fermion one, as
shown in Ref.~\cite{delAguila:2003bh}.  The latter one coincides with
that of a scalar with $\hat{m}_{i}=0$ and arbitrary brane kinetic
coefficients $r_{i}$~\cite{Carena:2004zn}.  Note that although
$\Phi_{2}$ obeys Dirichlet boundary conditions on both branes (hence
no brane mass terms for $\Phi_{2}$ are allowed), the singular terms of
Eq.~(\ref{SingularBKTs_OddScalar}) were not taken into account in the
derivation of the effective potential due to real scalars above.  In
particular, the contribution to $V_{\rm eff}$ from $\Phi_{2}$ does not
simply correspond to the limit $\hat{m}_{i} \to \infty$ of
Eq.~(\ref{FinalPot}) (except when $\hat{r}_{i}=0$).  However, since
the one-loop effective potential, Eq.~(\ref{EffPot}), depends only on
the KK spectrum, we can immediately write down the contribution from
$\Phi_{2}$ as $2 \times \left.  V_{\rm eff}(L)
\right|_{\hat{m}_{i}=0}$, where $V_{\rm eff}(L)$ is given in
Eq.~(\ref{FinalPot}).  Therefore, we can generalize
Eq.~(\ref{Veff_Hyper_nori}) to
\beqa
V^{\rm Hyper}_{\rm eff} &=& 2 \times \left. \rule{0mm}{5mm} V_{\rm eff}(L) \right|_{\alpha;\hat{r}_{i};\hat{m}_{i}} - 2 \times \left. \rule{0mm}{5mm} V_{\rm eff}(L) \right|_{\alpha;\hat{r}_{i};\hat{m}_{i}=0}~,
\label{Veff_Hyper}
\eeqa
where the contribution from the ``odd'' scalar precisely cancels half
of the fermion contribution.

For the 5D gauge supermultiplet, consisting of a vector superfield
$(A_{\mu}, \lambda_{1})$ with index $\alpha = 1$, and a chiral
superfield $(\Sigma+i A_{5},\lambda_{2})$ with index $\alpha=0$, one
has
\beqa
V^{\rm Vector}_{\rm eff} &=& 4 \times \left. \rule{0mm}{5mm} V_{\rm eff}(L) \right|_{\alpha=1;\hat{r}_{i};\hat{m}_{i}=0} - 4 \times \left. \rule{0mm}{5mm} V_{\rm eff}(L) \right|_{\alpha=1;\hat{r}_{i};\hat{m}_{i}}~,
\label{Veff_Vector}
\eeqa
where the second term comes from the gauginos (here $\hat{m}_{i}$
denote brane localized Majorana masses), while the first one arises
from $A_{M}$ (including $A_{5}$ and the associated Faddeev-Popov ghost
contributions) plus the term arising from the odd (real) scalar
$\Sigma$.  In the presence of BKT's, the action for the latter field
(after integrating out the auxiliary components) contains singular
terms of the form (\ref{SingularBKTs_OddScalar}) that ensure that the
KK spectrum of the $\Sigma$ field is identical to the gaugino KK
spectrum with $\hat{m}_{i} = 0$ (see Ref.~\cite{delAguila:2003bh} for
details), and its contribution is therefore identical to the gauge
boson one for any $\hat{r}_{i}$.

The 5D supergravity multiplet has the following propagating degrees of
freedom: the f\"unfbein $e^{A}_{M}$, a (Dirac) gravitino $\Psi_{M}$,
and a vector field $B_{M}$ (the graviphoton)~\cite{D'Auria:1981kq}
(see also Ref.~\cite{Altendorfer:2000rr}).  The fields containing
zero-modes are $(e_{\mu}^{a}, e_{5}^{\hat{5}}, B_{5}, \psi^{+}_{\mu},
\psi^{-}_{5})$, and their KK wavefunctions have index $\alpha = 2$.
The remaining fields, $(e_{5}^{a}, e_{\mu}^{\hat{5}}, B_{\mu},
\psi^{-}_{\mu}, \psi^{+}_{5})$, obey Dirichlet boundary conditions and
have index $\alpha=1$.  Here $\hat{5}$ denotes the fifth tangent space
index, and $\psi_{M}^{\pm} = \frac{1}{\sqrt{2}} \left( \psi^{1}_{M}
\pm \psi^{2}_{M} \right)$ where $\psi^{i}_{M}$ are the two Weyl
components of $\Psi_{M}$.  At each KK level $n\neq 0$, the KK
gravitons, $g_{\mu\nu}^{n}$, and KK graviphotons, $B_{\mu}^{n}$, gain
mass by eating $g_{\mu5}^{n}$, $g_{55}^{n}$ and $B_{5}^{n}$, while the
KK gravitinos $\Psi_{\mu}^{n}$ gain mass by eating
$\Psi_{5}^{n}$~\cite{Bagger:2001ep,DeCurtis:2003hs}.  Thus, the
effective potential from the gravity supermultiplet is
\beqa
V^{\rm SUGRA}_{\rm eff} &=& 8 \times \left. \rule{0mm}{5mm} V_{\rm eff}(L) \right|_{\alpha=2;\hat{r}_{i};\hat{m}_{i}=0} - 8 \times \left. \rule{0mm}{5mm} V_{\rm eff}(L) \right|_{\alpha=2;\hat{r}_{i};\hat{m}_{i}}~,
\label{Veff_Gravity}
\eeqa
where the second term corresponds to the gravitino contribution, while
the first term includes both the KK towers of 4D massive gravitons and
4D massive graviphotons (as in the cases discussed above, the
$B^{n}_{\mu}$, which obey Dirichlet boundary conditions and have
$\alpha=1$, contribute like a $(+,+)$ field with $\alpha=2$ and
vanishing brane mass terms).

Finally, we note that the radion multiplet, which contains the radion
scalar itself, as well as $B_5^0$, and $\Psi_5^0$, consists, before
stabilization, of massless zero modes.  As we are mainly interested in
stabilization scenarios based on bulk field quantum effects, we take
the radion multiplet to be massless, at leading order.  Therefore, it
does not contribute to the 1-loop potential computed above, which only
accounts for the effects of massive KK modes.  The contribution of the
radion multiplet to the radion scalar potential are then higher order
in our treatment and hence ignored.  With these results at hand, we
turn to the question of radiative radion stabilization in SUSY
theories in the next section.

\section{Application to SUSY Theories in RS}
\label{sec:susy}

We now consider an application to supersymmetric theories that are
softly broken by boundary masses.  As we have seen in the previous
section, the contribution to the radion potential from a pair of
superpartners (that have identical brane-kinetic terms, but may have
different boundary masses) is proportional to $\Delta V_{\rm
eff}(L;\hat{r}_{i},\hat{m}_{i}) = \left.  V_{\rm
eff}(L)\right|_{\hat{r}_{i},\hat{m}_{i}=0} - \left.  V_{\rm eff}(L)
\right|_{\hat{r}_{i},\hat{m}_{i}}$,~\footnote{\label{signconvention}The
sign corresponds to a gauge-gaugino (or gravity-gravitino) multiplet.
For a scalar-fermion multiplet the sign is opposite, since the soft
mass affects the scalar, which contributes with a plus sign to the
radion potential.} where $V_{\rm eff}(L)$ is given in
Eq.~(\ref{FinalPot}).  This takes the form
\beqa
\Delta V_{\rm eff}(L;\hat{r}_{i},\hat{m}_{i}) &=&  
\frac{\Lambda^2 k^{2} e^{-4kL}}{16\pi^{2}} u \, \hat{m}_{IR} +  
\Delta V_{\rm Casimir}(L) + \textrm{const.}
\label{DeltaVeff}
\eeqa
Here we included a quadratically divergent contribution,\footnote{The
quadratic divergence is proportional to the supertrace of the square
mass matrix, ${\rm sTr}{\cal M}^{2}$, which is non-vanishing in the
observable sector~\cite{Choi:1994xg}.  If the model in question
satisfies ${\rm sTr}{\cal M}^{2}$ = 0, then only a logarithmic
sensitivity to the cutoff remains~\cite{Intriligator:2007cp}.} where
one can estimate that the cutoff $\Lambda \lesssim 4\pi k$, with $u$
an incalculable dimensionless coefficient that might be expected to be
of order one [we absorb here the term proportional to ${\cal I}_{IR}$
in Eq.~(\ref{FinalPot})].  The first term can be expected to dominate
over $\Delta V_{\rm Casimir}(L)$ unless either $\Lambda \sim k$, or
one is willing to fine-tune $u$ to small values.  However, notice that
when $\hat{m}_{IR} = 0$, the first term in Eq.~(\ref{DeltaVeff})
vanishes, and the radion potential is cutoff-independent at one-loop
order.  We consider three different scenarios: \\

\begin{figure}[t]
\begin{center}
\includegraphics[width=0.5 \textwidth]{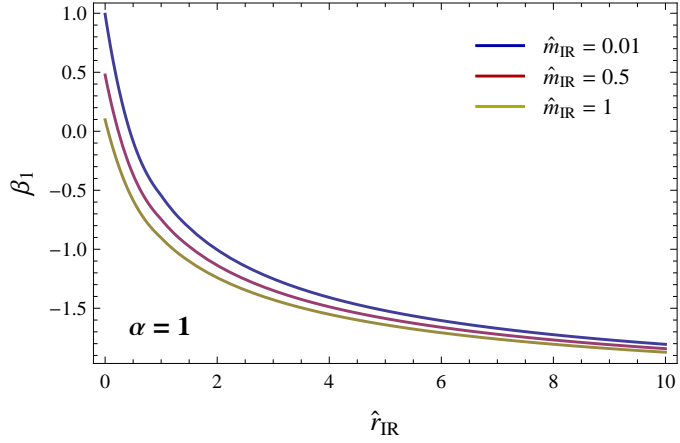}
\caption{ 
$\beta_{1} \equiv \int_{0}^{\infty} \! dt \, t^{3} \left[ \tilde{K}^{IR}_{\alpha=1}(t)/\tilde{I}^{IR}_{\alpha=1}(t) \right]$ as a function of $\hat{r}_{IR}$ for three values of $\hat{m}_{IR}$. 
}
\label{fig:beta1vsrIR}
\end{center}
\end{figure}
$\bullet$ \textit{High-scale SUSY breaking on the UV brane}: We
consider a region where $L > L_{T}$, where $L_{T}$ is defined in
Eq.~(\ref{LT}), so that Eq.~(\ref{VCasLargemUV}) applies for the
superpartner with a non-vanishing UV mass term.  Taking for
concreteness the case $\alpha = 1$, so that
Eq.~(\ref{VCasSmallmUValpha1}) applies to the superpartner with no
mass terms (e.g. the gauge field), we see that
\beqa 
\Delta V_{\rm Casimir}(L) &\approx& \left. V_{\rm Casimir}(L) \right|_{\hat{r}_{i},\hat{m}_{i} = 0}
~\approx~ - \frac{k^{4} e^{-4kL}}{16\pi^{2}} \, \frac{1}{k L + \hat{r}_{UV}}
\int_{0}^{\infty} \! dt \, t^{3} \frac{\tilde{K}^{IR}_{\alpha=1}(t)}{\tilde{I}^{IR}_{\alpha=1}(t)}~,
\label{Casimir_App1}
\eeqa
since the term $\left.  -V_{\rm Casimir}(L)
\right|_{\hat{r}_{i},\hat{m}_{i}}$ gives an exponentially small
contribution (suppressed by $e^{-2kL}$, as remarked in the paragraph
after Eq.~(\ref{VCasSmallmUV1alpha2}) of
Subsection~\ref{CasimirEnergy}).  This is essentially the result
derived in Ref.~\cite{Garriga:2002vf}, although here we allow for a
non-vanishing $\hat{r}_{UV}$.  If SUSY is unbroken on the IR brane,
the first term in Eq.~(\ref{DeltaVeff}) vanishes, and the radion is
not stabilized.  Depending on the size of the IR BKT, which can change
the sign of the integral in Eq.~(\ref{Casimir_App1}), the branes can
collapse towards each other or ``run-away''.

However, if SUSY is also broken on the IR brane, one can realize the
scenario envisioned in Ref.~\cite{Garriga:2002vf}.  In this case, the
radion potential, Eq.~(\ref{FinalPot}), has an extremum at $kL \approx
\beta_{1}/\Delta - \hat{r}_{UV}$, provided $\beta_{1}/\Delta > 0$,
where $\Delta = (\Lambda/k)^{2} u \, \hat{m}_{IR}$, and $\beta_{1}
\equiv \int_{0}^{\infty} \!  dt \, t^{3} \left[
\tilde{K}^{IR}_{\alpha=1}(t)/\tilde{I}^{IR}_{\alpha=1}(t) \right]$.
This extremum is a minimum when both $\beta_{1}$ and $\Delta$ are
negative.  The calculable coefficient $\beta_{1}$ can indeed be
negative if there is a non-negligible $\hat{r}_{IR}$ or a sizable
$\hat{m}_{IR}$, as shown in Fig.~\ref{fig:beta1vsrIR}.  Obtaining a
minimum with $kL \sim 30$ requires that $|\Delta| \sim 1/30$, which
can arise from a suppression due to $\hat{m}_{IR} \ll 1$.  Note that
having $\hat{m}_{IR} \ll 1$ does not mean that the superpartner ({\it
e.g} the gaugino) is light, since we are assuming that there is a
``large'' source of SUSY breaking on the UV brane (more precisely,
$e^{2kL}\hat{m}_{UV} \gg 1$).  In fact, Table.~\ref{Zero-mode_masses}
of Appendix~\ref{app:scalars} indicates that the physical soft
breaking mass is of order the KK scale, $M_{\rm KK} \sim 2\times
k\,e^{-kL}$, although there is a region where the soft mass can be
parametrically smaller, of order $k\,e^{-kL}/\sqrt{kL}$.  \\

$\bullet$ \textit{Low-scale SUSY breaking from the UV brane}: We now
consider SUSY breaking on the UV brane only ({\it i.e.} $\hat{m}_{IR}
= 0$, so that the incalculable contribution to the radion potential
vanishes), and focus on the case that $kL \gg 1$ with $L \ll L_{T}$,
as defined in Eq.~(\ref{LT}) ($\hat{m}_{UV}$ so small that there is a
light state in the spectrum).  As remarked in the discussion after
Eq.~(\ref{VCasSmallmUV1alpha2}), the leading order contributions
cancel within a given supermultiplet.  We can therefore immediately
obtain the leading effect in $\hat{m}_{UV}$ from
Eqs.~(\ref{UVRatioSmallmUV}) and (\ref{VCasimirApproxSmallmUV}) of
Appendix~\ref{app:AnalyticApprox}, for the various possible
localizations of the would-be zero mode.  For instance, for IR
localization, $0 < \alpha < 1$:
\beqa
\Delta V_{\rm eff}(L;\hat{r}_{i},\hat{m}_{UV},\hat{m}_{IR} = 0) &\approx& - \frac{k^{4} e^{-2(2-\alpha)kL}}{16\pi^{2}} \, \frac{4^{\alpha} \hat{m}_{UV} \sin^{2}(\pi\alpha) \Gamma(\alpha)^{2}}{\pi^{2}}
\int_{0}^{\infty} \! dt \, t^{3-2\alpha} \frac{\tilde{K}^{IR}_{\alpha}(t)}{\tilde{I}^{IR}_{\alpha}(t)}~,
\label{VCasSmallmUV0alpha1SUSY}
\eeqa
for the flat case, $\alpha = 1$:
\beqa
\Delta V_{\rm eff}(L;\hat{r}_{i},\hat{m}_{UV},\hat{m}_{IR} = 0) &\approx& - \frac{k^{4} e^{-2kL}}{16\pi^{2}} \, \frac{\hat{m}_{UV}}{(k L + \hat{r}_{UV})^{2}}
\int_{0}^{\infty} \! dt \, t \, \frac{\tilde{K}^{IR}_{\alpha=1}(t)}{\tilde{I}^{IR}_{\alpha=1}(t)}~,
\label{VCasSmallmUValpha1SUSY}
\eeqa
and for ``moderate'' UV localization, $1 < \alpha < 2$:
\beqa
\Delta V_{\rm eff}(L;\hat{r}_{i},\hat{m}_{UV},\hat{m}_{IR} = 0) &\approx& 
\nonumber \\ [0.5em]
& & \hspace{-5cm}
- \frac{k^{4} e^{-2\alpha kL}}{16\pi^{2}} \, \frac{4^{2 - \alpha} (\alpha - 1) \hat{m}_{UV}}{\Gamma(\alpha) \left[ \Gamma(\alpha - 1) + 2\hat{r}_{UV} \Gamma(\alpha) \right] [1 + 2 \hat{r}_{UV} (\alpha - 1)]}
\int_{0}^{\infty} \! dt \, t^{2\alpha-1} \frac{\tilde{K}^{IR}_{\alpha}(t)}{\tilde{I}^{IR}_{\alpha}(t)}~.
\label{VCasSmallmUV1alpha2SUSY}
\eeqa
For $\alpha \geq 2$, one also obtains $\Delta V_{\rm
eff}(L;\hat{r}_{i},\hat{m}_{UV},\hat{m}_{IR} = 0) \sim e^{-2\alpha
kL}$.  Note that the warp factor dependence is the same for $\alpha =
1 + \delta\alpha$ and $\alpha = 1 - \delta\alpha$, for any
$\delta\alpha$, although the coefficients do not exhibit this
symmetry.  In the above formulas it is understood that $\hat{m}_{IR} =
0$ in $\tilde{K}^{IR}_{\alpha}$ and $\tilde{I}^{IR}_{\alpha}$.  It is
clear that these potentials do not have a minimum by themselves.
Nevertheless, note that from the point of view of the warp factor
dependence, the case $\alpha = 1$ gives the largest contribution.
Furthermore, notice the ``double-volume'' suppression in
Eq.~(\ref{VCasSmallmUValpha1SUSY}).  These properties are important
when there are other quantum mechanical sources of radion
stabilization~\cite{HDEP}.  We also point out that the coefficient of
Eq.~(\ref{VCasSmallmUValpha1SUSY}) can have either sign, since for
instance
\beqa
\int_{0}^\infty \! dt \, t \, \frac{\tilde{K}^{IR}_{\alpha = 1}(t)}{\tilde{I}^{IR}_{\alpha = 1}(t)} &\longrightarrow&
\left\{
\begin{array}{lcl}
+ 0.63  & & \textrm{for } \hat{r}_{IR}=0   \\[0.5em]
- 0.10  & & \textrm{for } \hat{r}_{IR}=1
\end{array}
\right.~.
\eeqa
We now turn to the last scenario we wish to consider:
\\

$\bullet$ \textit{SUSY breaking from the IR brane only}: When
$\hat{m}_{UV} = 0$, but $\hat{m}_{IR} \ll 1$, the effective potential
is determined by the term linear in $\hat{m}_{IR}$ of
Eqs.~(\ref{VCasSmallmUV0alpha1}), (\ref{VCasSmallmUValpha1}) and
(\ref{VCasSmallmUV1alpha2}).  For instance, in the case of a vector multiplet with $\alpha = 1$, and for $kL \gg 1$, we have
\beqa
\Delta V_{\rm eff}(L;\hat{r}_{i},\hat{m}_{UV} = 0,\hat{m}_{IR}) &\approx&
\frac{k^{4} e^{-4kL}}{16\pi^{2}} \left[ \frac{\Lambda^{2}}{k^{2}} u \, \hat{m}_{IR} - \frac{B_{IR}\, \hat{m}_{IR} }{kL + \hat{r}_{UV}} \right]~,
\label{VeffmIRalpha1SUSY}
\eeqa
where [we set $\hat{m}_{IR} = 0$ in $\tilde{I}^{IR}_{1}(t)^{2}$]
\beqa
B_{IR} &=& \int_{0}^\infty \! dt \, \frac{t^{3}}{\tilde{I}^{IR}_{1}(t)^{2}} ~\longrightarrow ~
\left\{
\begin{array}{ccc}
1.26  & \textrm{for } \hat{r}_{IR}=0   \\[0.5em]
0.48  & \textrm{for } \hat{r}_{IR}=1
\end{array}
\right.
\eeqa
is manifestly positive.  In fact, for arbitrary $\hat{m}_{IR}$ the
Casimir energy part of the potential (i.e. the \textit{difference}
between Eq.~(\ref{VCasSmallmUValpha1}) at $\hat{m}_{IR}=0$ and the
same expression at arbitrary $\hat{m}_{IR}$) is always negative.
Thus, if $u < 0$ this potential does not have an extremum in the
physical (positive) region for $L$.  However, if $u > 0$ the effective
potential exhibits a \textit{maximum}.  It follows that the
contribution due to matter hypermultiplets with $\alpha = 1$ (which
has an overall minus sign relative to the gauge case, see
footnote~\ref{signconvention}) has a minimum.  Hence, a model with more
matter degrees of freedom (with $\alpha \approx 1$) than gauge degrees
of freedom can stabilize the size of the extra dimension through these
quantum effects.  However, for this minimum to be at $kL \gg 1$ (for
self-consistency with the above approximations), one must have $u
\Lambda^{2}/k^{2} \ll 1$.  For small $kL$, where fields with different
localization parameters $\alpha$ can give comparable contributions to
the radion potential, stabilization may be possible as in the flat
$S^{1}/Z_{2}$ orbifold case~\cite{Ponton:2001hq}.

\section{Conclusions}
\label{sec:conclusions}

In this work, we examined the use of supersymmetry to control vacuum
energy contributions to the radion potential.  Our analysis included
UV- and IR-brane-localized mass terms, to implement soft supersymmetry
breaking, as well as localized kinetic terms.  We found that UV-brane
breaking of supersymmetry leads to a diminished Casimir energy
contribution to the radion potential, both through the size of the
soft terms, as well as through an additional volume suppression.  This
finding may be useful in constructing models where dynamical EWSB
induces a non-trivial radion potential that is protected from unwanted
large corrections, thus tying the Planck-weak scale hierarchy to the
dynamics of EWSB itself~\cite{HDEP}.  As another possibility, we
studied the question of quantum radion stabilization by bulk fields
(without VEV's).  Stabilizing the radion at $kL \gg 1$ through the
quantum effects studied in this paper requires that SUSY be broken on
the IR brane.  If in addition SUSY is also broken on the UV brane, the
radion stabilization mechanism by flat fields of
Ref.~\cite{Garriga:2002vf} can be easily realized, provided the
\textit{sign} of an incalculable contribution is favorable.  Our
results are sufficiently general to be applicable to fields of
arbitrary spin, and obeying arbitrary boundary conditions, and can be
expected to be useful also outside the SUSY context.

\vspace{5mm} 
\noindent 
%

\noindent \textbf{Note added:} After posting the initial version of
this work, we found that some contributions were missing in our
brane-tension renormalization.  These contributions signal that,
absent exact supersymmetry, quantum corrections to brane-tensions are
not well-defined in our effective theory, as may also be inferred on
general grounds.  We have clarified this point in the new version and
removed discussions dealing with brane-tension renormalization, since
they do not affect our main conclusions outlined in the abstract.

\vspace{5mm} 
\noindent 
%

\subsection*{Acknowledgements}
We thank Markus Luty for useful discussions.  The work of H.D. is
supported by the United States Department of Energy under Grant
Contract DE-AC02-98CH10886.  E.P. is supported by DOE under contract
DE-FG02-92ER-40699.

\appendix

\section{Scalar Fields with Brane-Localized Terms}
\label{app:scalars}

In this appendix we collect, for reference, various details of the KK
decomposition of a scalar field with arbitrary localized terms.  We
also give approximate analytical expressions for the lightest KK mass
and the corresponding KK eigenfunction.

\subsection{KK Decomposition}
\label{app:KKDecomposition}

The action of Eq~(\ref{Action}) for the scalar field, $\Phi$, is
supplemented by the boundary conditions
\beqa
\left. (\partial_{y} \Phi - M_{0} \Phi - r_{UV} \Box \Phi)  \right|_{y=0} = 0~,
\hspace{1cm}
\left. (\partial_{y} \Phi + M_{L} \Phi + e^{2kL} r_{IR} \Box \Phi)  \right|_{y=L} = 0~,
\label{bcs}
\eeqa
where $\Box = \eta^{\mu\nu} \partial_{\mu} \partial_{\nu}$.  We write
the KK decomposition for $\Phi$ as
\beqa
\Phi(x^{\mu},y) = \frac{e^{ky}}{\sqrt{L}} \sum^{\infty}_{n=0} \phi^{n}(x^{\mu}) f_{n}(y)~,
\label{KKPhi}
\eeqa
where we pulled out an explicit factor $e^{ky}$ for convenience.  The
$\phi^{n}$ obey the Klein-Gordon equation $(\Box + m^{2}_{n} )\phi^{n}
= 0$, while the KK wavefunctions satisfy
\beqa
f''_{n} - 2k f'_{n} - (3k^{2} + M^{2}) f_{n} =
- e^{2ky} m^{2}_{n} f_{n}~,
\label{EQf}
\eeqa
and are explicitly given by
\beqa
f_{n}(y) = A_{n} e^{ky} \left[ J_{\alpha} \!
\left(\frac{m_{n}}{k} e^{ky} \right)
+ B_{n} \, Y_{\alpha} \! \left(\frac{m_{n}}{k} e^{ky} \right)
\right]~,
\label{fn}
\eeqa
where $\alpha$ was defined in Eqs.~(\ref{Mdef}) and (\ref{mLdef}), and
$A_{n}$ is a normalization constant, determined from
\beqa
\frac{1}{L} \int^{L}_{0} \! dy f_{n}(y) f_{m}(y) = \delta_{nm}~.
\label{KKnormalization}
\eeqa
In terms of the definitions (\ref{mLdef}) and (\ref{bUVIR}), the
boundary conditions of Eq.~(\ref{bcs}), become
\beqa
\partial_{y} f_{n} |_{y = 0} &=& - \left[ \left(\alpha - 1 \right) + b_{UV}\left(\frac{m_{n}}{k}\right) \right] k f_{n}(0)~,
\label{BC0}
\\ [0.5em]
\partial_{y} f_{n} |_{y = L} &=& - \left[ \left(\alpha - 1 \right) - b_{IR}\left(\frac{m_{n}}{k} e^{kL}\right) \right] k f_{n}(L)~,
\label{BCL}
\eeqa
which determine the constant $B_{n}$ in Eq.~(\ref{fn}) and the
spectrum $m_{n}$ according to
\beqa
B_{n} = - \frac{\tilde{J}^{UV}_{\alpha} \! \left(\frac{m_{n}}{k} \right)}{\tilde{Y}^{UV}_{\alpha} \! \left(\frac{m_{n}}{k} \right)}
=  - \frac{\tilde{J}^{IR}_{\alpha} \! \left(\frac{m_{n}}{k} e^{kL} \right)}{\tilde{Y}^{IR}_{\alpha} \! \left(\frac{m_{n}}{k} e^{kL}  \right)}~,
\label{bsolns}
\eeqa
where we used the definitions (\ref{JUVtilde}) and (\ref{JIRtilde})
for $\tilde{J}^{UV}_{\alpha}$, $\tilde{J}^{IR}_{\alpha}$,
$\tilde{Y}^{UV}_{\alpha}$ and $\tilde{Y}^{IR}_{\alpha}$.

\subsection{Lightest KK Modes}
\label{app:LightStates}

The second equation in (\ref{bsolns}) determines the mass eigenvalues
$m_{n}$.  In particular, when $m_{UV} = m_{IR} = 0$, one finds that
$m_{0} = 0$ is a solution.  For non-zero but small $m_{UV}$ or
$m_{IR}$ there is an eigenstate with mass $m_{0} \ll k\,e^{-kL}$.  In
this limit we can also obtain the approximate profile for the light
mode by setting $m_{n} \approx 0$ in Eq.~(\ref{EQf}), which is then
solved by
\beqa
f_{0}(y) &\approx& \sqrt{ \frac{2kL(1-\alpha)}{e^{2(1-\alpha)kL} - 1} } \,
e^{(1-\alpha) k y}~,
\eeqa
where we also imposed the boundary conditions (\ref{BC0}) and
(\ref{BCL}) with $m_{UV} = m_{IR} = 0$, and normalized according to
Eq.~(\ref{KKnormalization}).  We see that this is the same profile as
that for a zero-mode fermion with a bulk Dirac mass $M = c k$, where $c=
\alpha - 1/2$, and in particular that the state is localized near the
UV (IR) brane for $c > 1/2$ ($c < 1/2$).  The case $c = 1/2$
corresponds to a flat profile, $f_{0}(y) = 1$.

\begin{table}[t]
\begin{center}
\setlength{\extrarowheight}{8pt}
\begin{tabular}{c||cc}
& $m_{UV}\neq 0$ and $m_{IR} = 0$ & $m_{UV} = 0$ and $m_{IR}\neq 0$  \\ [0.5em]
\hline
\hline
\rule{0mm}{1.1cm}
$\alpha > 1$ (i.e.~$c > 1/2$) & 
$\displaystyle \sqrt{\frac{4 \alpha e^{2kL} \hat{m}_{UV}}{e^{2kL} \hat{m}_{UV} + 2 \alpha/(\alpha-1)}}$
&
$\displaystyle \sqrt{\frac{4 \alpha (\alpha - 1)\hat{m}_{IR}}
{-\alpha (2+\hat{m}_{IR}) + (2 \alpha + \hat{m}_{IR})e^{2(\alpha-1)kL}}}$ \\ [1.5em]
\rule{0mm}{1.1cm}
$\alpha = 1$ (i.e.~$c = 1/2)$ &
$\displaystyle \sqrt{\frac{4 e^{2kL} \hat{m}_{UV}}{e^{2kL} \hat{m}_{UV} + 4kL}}$
&
$\displaystyle \sqrt{\frac{4 \hat{m}_{IR}}{2(2+\hat{m}_{IR})kL- \hat{m}_{IR}}}$ \\ [1.5em]
\rule{0mm}{1.1cm}
$\alpha < 1$ (i.e.~$c < 1/2$)&
$\displaystyle \sqrt{\frac{4 \alpha e^{2 \alpha kL} \hat{m}_{UV}}{e^{2 \alpha kL} \hat{m}_{UV} - 2 \alpha/(\alpha-1)}}$
&
$\displaystyle \sqrt{\frac{4 (1-\alpha) \hat{m}_{IR}}{2+\hat{m}_{IR}}}$ \\ [1.5em]
\end{tabular}
\end{center}
\caption{Would-be zero mode mass, $m_{0}$, in units of $k \, e^{-kL}$.}
\label{Zero-mode_masses}
\end{table}%

We can obtain an analytical approximation to the mass $m_0$ when
$m_{UV}$ or $m_{IR}$ are small, by using the small argument
approximation for the Bessel functions.  In
Table~\ref{Zero-mode_masses}, we summarize the results for $m_{0}/(k
\, e^{-kL})$ for the cases $\alpha > 1$ (i.e. $c > 1/2$), $\alpha = 1$
(i.e. $c = 1/2$) and $\alpha < 1$ (i.e. $c < 1/2$) when one of the UV
or IR brane-localized masses is non-vanishing, but the other one
vanishes.  For simplicity, we set the brane kinetic terms to zero.  We
see that for non-vanishing $m_{UV}$, the criterion for these solutions
to be valid (that they be small compared to $k \, e^{-kL}$) requires
that $e^{2kL} \hat{m}_{UV}$ be small when $\alpha \geq 1$, while for
$\alpha < 1$ one needs $e^{2 \alpha kL} \hat{m}_{UV} \ll 1$.  This is
related to the transition length $L_{T}$ defined in Eq.~(\ref{LT}).
Note also that when $m_{IR} \neq 0$ and $\alpha > 1$, $m_0$ is
exponentially small compared to $k \, e^{-kL}$, as long as $kL\gg 1$,
due to the UV-localization of the corresponding wavefunction.

\section{Simple Expressions for the Radion Potential}
\label{app:AnalyticApprox}

In this appendix, we derive an analytic approximation to the radion
potential that is sufficiently accurate for most applications and
exhibits the radion dependence explicitly.  We focus on the Casimir
energy piece, Eq.~(\ref{CasimirPot}), that we reproduce here for easy
reference:
\beqa
V_{\rm Casimir}(L) &=& \frac{k^{4} e^{-4kL}}{16\pi^{2}} \int_{0}^{\infty} \! dt \, t^{3} \ln \left| 1 - \frac{\tilde{K}^{IR}_{\alpha}(t)}{\tilde{I}^{IR}_{\alpha}(t)} \frac{\tilde{I}^{UV}_{\alpha}(t e^{-kL})}{\tilde{K}^{UV}_{\alpha}(t e^{-kL})} \right|~.
\label{CasimirPotAppendix}
\eeqa
Note that the integral receives the main contribution from the
region $0 < t < {\rm few}$ as a consequence of the asymptotic
behavior of the ``IR factor''
\beqa
\frac{\tilde{K}^{IR}_{\alpha}(t)}{\tilde{I}^{IR}_{\alpha}(t)} &\sim& - \frac{\hat{m}_{IR} - t + \hat{r}_{IR} t^{2}}{\hat{m}_{IR} + t + \hat{r}_{IR} t^{2}} \, \pi \, e^{-2t}~,
\hspace{1cm} \textrm{as } t \to \infty~,
\eeqa
which implies that the integrand in Eq.~(\ref{CasimirPotAppendix})
tends to zero exponentially for $t > {\rm few}$.  It follows that, for
$e^{-k L} \ll 1$, we can approximate the ``UV factor'' in
Eq.~(\ref{CasimirPotAppendix}) --which contains the radion
dependence-- by~\footnote{We only use the leading order small argument
approximations for the modified Bessel functions:
\beqa
I_{n}(x) &\approx& \frac{1}{\Gamma(n + 1)} \left( \frac{x}{2} \right)^{n}~,
\hspace{1cm}
K_{n}(x) ~\approx~ 
\left\{
\begin{array}{lcl}
-\gamma_{E} - \ln \left( x/2 \right)~,
&   &  \textrm{for } n = 0~, \\ [0.5em]
\frac{1}{2} \left[ \Gamma(n) \left( \frac{x}{2} \right)^{-n} +\Gamma(-n) \left( \frac{x}{2} \right)^{n} \right]~,
&   & \textrm{for } 0 < |n| < 1~,  \\ [0.5em]
\frac{1}{2} \, \Gamma(|n|) \left( \frac{x}{2} \right)^{-|n|}~,
&   &  \textrm{for } |n| \geq 1~.
\end{array}
\right.
\nonumber
\eeqa
}
\beqa
\frac{\tilde{I}^{UV}_{\alpha}(t e^{-kL})}{\tilde{K}^{UV}_{\alpha}(t e^{-kL})} &\approx&
4 \left[ \Xi(\alpha) - \frac{\Gamma(\alpha + 1)}{\left( \frac{t}{2} \, e^{-k L}  \right)^{2\alpha} } \, \frac{\Theta(\alpha) \Gamma(\alpha - 1) t^{2} + 2 \Gamma(\alpha) \left( \hat{r}_{UV} t^{2} + e^{2kL} \hat{m}_{UV} \right)}{\hat{r}_{UV} t^{2} + e^{2kL} (\hat{m}_{UV} - 2 \alpha)} \right]^{-1}~,
\label{UVFactorSmallArgGeneral}
\eeqa
where $\Theta(\alpha)$ is the Heaviside step function (unity for
$\alpha > 0$, vanishing for $\alpha < 0$), and
\beqa
\Xi(\alpha) &\equiv& 
\left\{
\begin{array}{lcl}
\displaystyle
2\pi \csc(\pi \alpha)  
&   & \textrm{for } 0 < |\alpha| < 1  \\ [0.5em]
\displaystyle
-\frac{4 \pi e^{2 kL} \alpha \csc(\pi \alpha)}{
\hat{r}_{UV} t^2 + e^{2 kL} \left( \hat{m}_{UV} - 2 \alpha \right)}  
&   & \textrm{for } 1 < \alpha < 2  \\ [1.1em]
\displaystyle
0  &   & \textrm{for } \alpha \geq 2
\end{array}
\right.~.
\eeqa
The cases $\alpha = 0$ and $\alpha = 1$ (and their very close
vicinity) require special treatment:
\beqa
\frac{\tilde{I}^{UV}_{\alpha = 0}(t e^{-kL})}{\tilde{K}^{UV}_{\alpha = 0}(t e^{-kL})} &\approx& 
\frac{(1/2 - \hat{r}_{UV}) \left(t \, e^{-kL} \right)^2 - \hat{m}_{UV}}{1 + \hat{m}_{UV} [ kL - \gamma_{E} - \ln(t/2)]}~,
\label{UVFactorSmallArgAlpha0}
\\ [0.5em]
\frac{\tilde{I}^{UV}_{\alpha = 1}(t e^{-kL})}{\tilde{K}^{UV}_{\alpha = 1}(t e^{-kL})} &\approx& 
- \frac{1}{2} \left(t \, e^{-kL} \right)^2 \frac{ \hat{r}_{UV} t^2 + e^{2 kL} (\hat{m}_{UV} - 2)}{ 
[kL + \hat{r}_{UV} - \gamma_{E} - \ln(t/2)] \, t^2 + e^{2 kL} \hat{m}_{UV}}~,
\label{UVFactorSmallArgAlpha1}
\eeqa
while for $\alpha \leq -1$, the ``UV factor'' is $\hat{r}_{UV}$ and
$\hat{m}_{UV}$-independent, and simply equals $(2/\pi) \sin(\pi
\alpha)$.  The cases were $\alpha$ is a negative integer, $-n$, should
be understood as the limit $\alpha \to -n$.
Eqs.~(\ref{UVFactorSmallArgGeneral})--(\ref{UVFactorSmallArgAlpha1})
can be used as a starting point for further approximations that hold
in different regimes of $L$.  In particular, taking into account that
$t$ is always of order one in the relevant region of integration,
these expressions indicate that there is a transition in the $L$
dependence of the potential according to whether $L \gg L_{\rm T}$ or
$L \ll L_{\rm T}$, where $L_{\rm T}$ is determined by $\hat{m}_{UV}$
according to Eq.~(\ref{LT}) of the main text.

For $L \gg L_{\rm T}$ (i.e. $e^{2kL} \hat{m}_{UV} \gg 1$), the second
term in the denominator of Eq.~(\ref{UVFactorSmallArgGeneral}) can be
approximated by $\left( \frac{t}{2} \, e^{-kL} \right)^{-2\alpha}$
$\left[2\hat{m}_{UV} \Gamma(\alpha) \Gamma(\alpha +
1)\right]/\left[\hat{m}_{UV} - 2\alpha \right]$.  It follows that for
$\alpha > 0$ this second term dominates over $\Xi(\alpha)$, while for
$\alpha < 0$ the $\Xi(\alpha)$ term dominates.  We then see that, for
$\alpha > 0$, the argument of the logarithm in
Eq.~(\ref{CasimirPotAppendix}) differs from unity by order
$e^{-2\alpha k L}/\hat{m}_{UV} \ll 1$, and we can also safely Taylor
expand it.  Therefore, we can write for $L \gg L_{\rm T}$:
\beqa
V^{\alpha > 0}_{\rm Casimir} &\approx& \frac{k^{4} e^{-2(2+\alpha)kL}}{16\pi^{2}} \frac{2(\hat{m}_{UV} - 2 \alpha) }{2^{2\alpha}\hat{m}_{UV} \Gamma(\alpha) \Gamma(\alpha + 1)}
\int_{0}^{\infty} \! dt \, t^{3+2\alpha} \frac{\tilde{K}^{IR}_{\alpha}(t)}{\tilde{I}^{IR}_{\alpha}(t)}~,
\label{VCasimirApproxLargemUV}
\eeqa
which also holds for $\alpha = 1$, while for $\alpha = 0$ one can use
\beqa
V^{\alpha = 0}_{\rm Casimir} &\approx& \frac{k^{4} e^{-4kL}}{16\pi^{2}} \frac{\hat{m}_{UV}}{1 + \hat{m}_{UV} (kL - \gamma_{E})}
\int_{0}^{\infty} \! dt \, t^{3} \frac{\tilde{K}^{IR}_{\alpha=0}(t)}{\tilde{I}^{IR}_{\alpha=0}(t)}~.
\label{VCasimirApproxLargemUValpha0}
\eeqa
Here, for simplicity, we neglected the $\ln(t/2)$ term in
Eq.~(\ref{UVFactorSmallArgAlpha0}), which introduces a modest error
(it is easy to keep it under the integral for numerical evaluation, if
necessary).  For negative index, when the $\Xi(\alpha)$ term in
Eq.~(\ref{UVFactorSmallArgGeneral}) dominates, we have instead
\beqa
V^{\alpha < 0}_{\rm Casimir} &\approx& \frac{k^{4} e^{-4kL}}{16\pi^{2}}
\int_{0}^{\infty} \! dt \, t^{3} \ln \left|1 - \frac{2 \sin(\pi\alpha)}{\pi} \, \frac{\tilde{K}^{IR}_{\alpha}(t)}{\tilde{I}^{IR}_{\alpha}(t)} \right|~.
\label{VCasLargemUVNegalpha}
\eeqa

In the complementary regime, $L \ll L_{T}$, we need to distinguish
several cases.  Keeping up to linear order in $\hat{m}_{UV}$, we have
that $\tilde{I}^{UV}_{\alpha}(t e^{-kL})/\tilde{K}^{UV}_{\alpha}(t
e^{-kL})$ can be approximated by
\beqa
\left\{
\begin{array}{lcl}
\displaystyle 
\frac{1}{2} e^{-2 kL} (1 - 2 \hat{r}_{UV}) t^2 - \hat{m}_{UV}~,  
&   & \textrm{for } \alpha = 0~,   \\ [1.5em]
\displaystyle 
\frac{2 \sin(\pi\alpha)}{\pi} \left[ 1 - \frac{\sin(\pi\alpha) \Gamma(\alpha)^{2}}{2\pi} \left( \frac{t}{2} \right)^{-2\alpha} e^{2\alpha kL} \hat{m}_{UV} \right]~,
&   &  \textrm{for } 0 < |\alpha| < 1~, \\ [1.5em]
\displaystyle
\frac{1}{k L + \hat{r}_{UV} - \gamma_{E} - \ln(t/2)} \left[ 1 - \frac{e^{2kL} \hat{m}_{UV} }{t^{2}\left( k L + \hat{r}_{UV} - \gamma_{E} - \ln(t/2) \right)} \right]~,
&   & \textrm{for } \alpha = 1~, \\ [1.5em]
\displaystyle
\frac{2 e^{-2(\alpha - 1)kL}}{ \Gamma(\alpha) \left[ \Gamma(\alpha - 1) + 2\hat{r}_{UV} \Gamma(\alpha) \right]}
\left( \frac{t}{2} \right)^{2(\alpha - 1)} \left[ 1 - \frac{2 (\alpha - 1) e^{2 kL} \hat{m}_{UV}}{t^2 [1 + 2 \hat{r}_{UV} (\alpha - 1)]} \right]~,
&   & \textrm{for } 1 < \alpha < 2~, \\ [1.5em]
\displaystyle
\frac{2 e^{-2(\alpha - 1)kL}}{ \Gamma(\alpha) \Gamma(\alpha - 1) \left[ 1 + 2\hat{r}_{UV} (\alpha -1) \right]}
\left( \frac{t}{2} \right)^{2(\alpha - 1)} \left[ 1 - \frac{2(\alpha -1) e^{2kL} \hat{m}_{UV}}{t^{2} \left[ 1 + 2 \hat{r}_{UV} (\alpha - 1) \right]} \right]~,
&   & \textrm{for } \alpha \geq 2~.
\end{array}
\right.
\label{UVRatioSmallmUV}
\eeqa
When $\alpha > 0$, these approximations assume that $t^2 \gg e^{2 kL}
\hat{m}_{UV}$ (for $\alpha \geq 1$) or $t^{2 \alpha} \gg e^{2 \alpha
kL} \hat{m}_{UV}$ (for $0 < \alpha < 1$), and therefore they fail for
very small $t$.  However, due to the $t^3$ factor in
Eq.~(\ref{CasimirPotAppendix}) the contribution to the integral from
the small $t$ region is negligible, and the above approximations can
be used everywhere.  One can see from the above expressions that the
``UV factor'' is very small compared to unity for $\alpha \geq 1$.
Therefore, one can expand the logarithm in
Eq.~(\ref{CasimirPotAppendix}) as was done in the large $L$ region, so
that for $L \ll L_{T}$ and $\alpha \geq 1$:
\beqa
V_{\rm Casimir}(L) &\approx& - \frac{k^{4} e^{-4kL}}{16\pi^{2}} \int_{0}^{\infty} \! dt \, t^{3} \frac{\tilde{K}^{IR}_{\alpha}(t)}{\tilde{I}^{IR}_{\alpha}(t)} \frac{\tilde{I}^{UV}_{\alpha}(t e^{-kL})}{\tilde{K}^{UV}_{\alpha}(t e^{-kL})}~,
\label{VCasimirApproxSmallmUV}
\eeqa
where $\tilde{I}^{UV}_{\alpha}(t e^{-kL})/\tilde{K}^{UV}_{\alpha}(t
e^{-kL})$ is replaced by the appropriate case in
Eq.~(\ref{UVRatioSmallmUV}).  In fact, this expression gives
reasonably accurate results even for $0 < \alpha < 1$.  For $\alpha <
0$, one can use Eq.~(\ref{VCasLargemUVNegalpha}) since the term linear
in $\hat{m}_{UV}$, which is non-vanishing only for $-1 < \alpha < 0$
[second line of Eq.~(\ref{UVRatioSmallmUV})], gives an exponentially
small correction.  Specific applications that can be easily derived
from the above expressions are further discussed in the main text.



\begin{thebibliography}{99}

\bibitem{RS}
  L.~Randall and R.~Sundrum,
  Phys.\ Rev.\ Lett.\  {\bf 83}, 3370 (1999)
  [arXiv:hep-ph/9905221].

\bibitem{Davoudiasl:1999tf}
  H.~Davoudiasl, J.~L.~Hewett and T.~G.~Rizzo,
  Phys.\ Lett.\  B {\bf 473}, 43 (2000)
  [arXiv:hep-ph/9911262];

\bibitem{Pomarol:1999ad}
  A.~Pomarol,
  Phys.\ Lett.\  B {\bf 486}, 153 (2000)
  [arXiv:hep-ph/9911294].

\bibitem{Grossman:1999ra}
  Y.~Grossman and M.~Neubert,
  Phys.\ Lett.\  B {\bf 474}, 361 (2000)
  [arXiv:hep-ph/9912408].


\bibitem{Gherghetta:2000qt}
  T.~Gherghetta and A.~Pomarol,
  Nucl.\ Phys.\  B {\bf 586} (2000) 141
  [arXiv:hep-ph/0003129].

\bibitem{Bai:2008gm}
  Y.~Bai, M.~Carena and E.~Pont\'on,
  Phys.\ Rev.\  D {\bf 81}, 065004 (2010)
  [arXiv:0809.1658 [hep-ph]].

\bibitem{Davoudiasl:2009xz}
  H.~Davoudiasl and E.~Pont\'on,
  Phys.\ Lett.\  B {\bf 680}, 247 (2009)
  [arXiv:0903.3410 [hep-ph]].

\bibitem{Davoudiasl:2002ua}
  H.~Davoudiasl, J.~L.~Hewett and T.~G.~Rizzo,
  Phys.\ Rev.\  D {\bf 68}, 045002 (2003)
  [arXiv:hep-ph/0212279].

\bibitem{Carena:2002dz}
  M.~S.~Carena, E.~Pont\'on, T.~M.~P.~Tait and C.~E.~M.~Wagner,
  Phys.\ Rev.\  D {\bf 67} (2003) 096006
  [arXiv:hep-ph/0212307];
  M.~S.~Carena, A.~Delgado, E.~Pont\'on, T.~M.~P.~Tait and C.~E.~M.~Wagner,
  Phys.\ Rev.\  D {\bf 68}, 035010 (2003)
  [arXiv:hep-ph/0305188].

\bibitem{Carena:2004zn}
  M.~S.~Carena, A.~Delgado, E.~Pont\'on {\it et al.},
  Phys.\ Rev.\  {\bf D71}, 015010 (2005).
  [hep-ph/0410344].
  
\bibitem{HDEP}
H. Davoudiasl and E. Pont\'{o}n, work in progress.  

\bibitem{Garriga:2000jb}
  J.~Garriga, O.~Pujolas and T.~Tanaka,
  Nucl.\ Phys.\  B {\bf 605}, 192 (2001)
  [arXiv:hep-th/0004109].

\bibitem{Goldberger:2000dv}
  W.~D.~Goldberger and I.~Z.~Rothstein,
  Phys.\ Lett.\  B {\bf 491}, 339 (2000)
  [arXiv:hep-th/0007065].

\bibitem{Garriga:2002vf}
  J.~Garriga and A.~Pomarol,
  Phys.\ Lett.\  B {\bf 560}, 91 (2003)
  [arXiv:hep-th/0212227].

\bibitem{Maru:2010ap}
 N.~Maru and Y.~Sakamura,
 JHEP {\bf 1004}, 100 (2010)
 [arXiv:1002.4259 [hep-ph]].

\bibitem{Goldberger:1999uk}
  W.~D.~Goldberger, M.~B.~Wise,
  Phys.\ Rev.\ Lett.\  {\bf 83}, 4922-4925 (1999).
  [hep-ph/9907447].

\bibitem{Flachi:2003bb}
  A.~Flachi, J.~Garriga, O.~Pujolas and T.~Tanaka,
  JHEP {\bf 0308}, 053 (2003)
  [arXiv:hep-th/0302017].

\bibitem{Gregoire:2004nn}
  T.~Gregoire, R.~Rattazzi, C.~A.~Scrucca, A.~Strumia and E.~Trincherini,
  Nucl.\ Phys.\  B {\bf 720}, 3 (2005)
  [arXiv:hep-th/0411216].

\bibitem{Katz:2006mva}
  A.~Katz, Y.~Shadmi and Y.~Shirman,
  Phys.\ Rev.\  D {\bf 75}, 055008 (2007)
  [arXiv:hep-th/0601036].
  
\bibitem{delAguila:2003bh}
  F.~del Aguila, M.~Perez-Victoria and J.~Santiago,
  JHEP {\bf 0302}, 051 (2003)
  [arXiv:hep-th/0302023].

\bibitem{ArkaniHamed:2001tb}
  N.~Arkani-Hamed, T.~Gregoire and J.~G.~Wacker,
  JHEP {\bf 0203}, 055 (2002)
  [arXiv:hep-th/0101233].

\bibitem{Marti:2001iw}
  D.~Marti and A.~Pomarol,
  Phys.\ Rev.\  D {\bf 64}, 105025 (2001)
  [arXiv:hep-th/0106256].

\bibitem{Mirabelli:1997aj}
  E.~A.~Mirabelli, M.~E.~Peskin,
  Phys.\ Rev.\  {\bf D58}, 065002 (1998).
  [hep-th/9712214].
  
\bibitem{D'Auria:1981kq}
  R.~D'Auria, E.~Maina, T.~Regge {\it et al.},
  Annals Phys.\  {\bf 135}, 237-269 (1981). 

\bibitem{Altendorfer:2000rr}
  R.~Altendorfer, J.~Bagger and D.~Nemeschansky,
  Phys.\ Rev.\  D {\bf 63}, 125025 (2001)
  [arXiv:hep-th/0003117].

\bibitem{Bagger:2001ep}
  J.~Bagger, F.~Feruglio and F.~Zwirner,
  JHEP {\bf 0202}, 010 (2002)
  [arXiv:hep-th/0108010].

\bibitem{DeCurtis:2003hs}
  S.~De Curtis, D.~Dominici, J.~R.~Pelaez,
  JHEP {\bf 0401}, 052 (2004).
  [hep-th/0311226].

\bibitem{Choi:1994xg}
  K.~Choi, J.~E.~Kim and H.~P.~Nilles,
  Phys.\ Rev.\ Lett.\  {\bf 73}, 1758 (1994)
  [arXiv:hep-ph/9404311].

\bibitem{Intriligator:2007cp}
  K.~A.~Intriligator and N.~Seiberg,
  Class.\ Quant.\ Grav.\  {\bf 24}, S741 (2007)
  [arXiv:hep-ph/0702069].

\bibitem{Ponton:2001hq}
  E.~Pont\'on, E.~Poppitz,
  JHEP {\bf 0106}, 019 (2001).
  [hep-ph/0105021].

\end{thebibliography}
\end{document}